\documentclass{elsart}
\usepackage{graphicx}
\usepackage{amssymb}

\newcommand{\upic}{$\mu$-PIC}
\newcommand{\microm}{\mathrm{\mu m}}

\def\Ltsim{\mathrel{\vbox to 0pt{\hbox{$\sim$}}\llap{$<$}}}  
\def\Gtsim{\mathrel{\vbox to 0pt{\hbox{$\sim$}}\llap{$>$}}}

\begin{document}
\intextsep 30pt
\textfloatsep 30pt
\begin{frontmatter}
\title{Simulation study of electron drift and gas multiplication in 
Micro Pixel Chamber}
\author[1]{\corauthref{cor1}Tsutomu~Nagayoshi}\ead{nagayosi@cr.scphys.kyoto-u.ac.jp},
\author[1]{Atsushi~Takada},
\author[1]{Hidetoshi~Kubo},
\author[1]{Kentaro~Miuchi},
\author[1]{Reiko~Orito},
\author[1]{Yoko~Okada},
\author[1]{Atsushi~Takeda},
\author[1]{Toru~Tanimori},
\author[1]{Masaru~Ueno},
\author[2]{Oleg~Bouianov},
\author[3]{Marina~Bouianov}
\address[1]{Cosmic-Ray Group, Department of Physics, Faculty of Science, Kyoto University Kitashirakawa, Sakyo-ku, Kyoto, 606-8502, Japan}
\address[2]{Laboratory for Nuclear Science, Massachusetts Institute of Technology, Cambridge, MA 02139, USA }
\address[3]{CSC-Scientific Computing Ltd., Espoo 02101, Finland}
\corauth[cor1]{Corresponding author.\\~~~Tel.$:$+81(0)75-753-3867$\;$fax$:$+81(0)75-753-3799}

\begin{abstract}

The physical processes of charge collection and gas multiplication of a Micro
Pixel Chamber ({\upic}) were studied in detail using a three-dimensional 
simulation. 
The collection efficiencies of primary electrons and gas multiplication
factors were calculated for several electrode structures. 
Based on those studies, we analyzed the optimization of the electrode structure
of the {\upic}, in order to obtain a high gas gain of more than $10^{4}$ 
and a simultaneous suppression of discharges. 
Consequently, we found that these characteristics strongly depend on 
the substrate thickness and the anode diameter of the {\upic}.
In addition, a gas gain of $10^{5}$ would be expected for a $\mu$-PIC having a 
thick substrate of $> 150\microm$.

\end{abstract}
\begin{keyword}
Gaseous detector; Micro-Pattern detector; Imaging; 3D simulation;
\end{keyword}
\end{frontmatter}

\section{Introduction}
\label{sec:Introduction}
\begin{figure}[hp]
\begin{center}
  \includegraphics[width=0.7\linewidth]{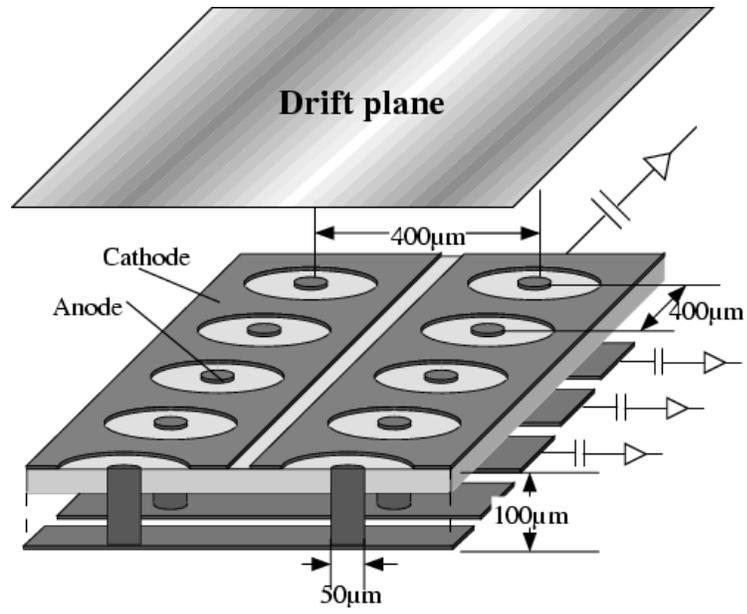}
  \caption{Schematic structure of the {\upic}. Anode pillars 
are formed through the substrate with a pitch of $400\microm$.}
  \label{fig:uPICstruct_bw.eps}
\end{center}
\end{figure}

The Micro Pixel Chamber ({\upic}) is a recently developed gaseous 
two-dimensional imaging detector~\cite{Ochi1}. 
The schematic structure is shown in Fig.~\ref{fig:uPICstruct_bw.eps}. 
The {\upic} is a double-sided Printed Circuit Board (PCB) with 
a $100\microm$ thick polyimide substrate. 
We developed a {\upic} with a $\mathrm{10 \times 10 cm^{2}}$ detection area. 
On the front and rear sides of the substrate,
256 cathode and 256 anode strips are formed orthogonally 
with a pitch of $400\microm$, respectively. 
A cathode strip has circular openings of $260\microm$ diameters 
with a pitch of $400\microm$. 
Anode pillars of $\sim$ 50$\microm$ diameters are formed on anode strips 
through the substrate at the center of each cathode openings. 
We have made several {\upic}s with various electrode structures 
and studied their performances~\cite{Nagayoshi1}.
Consequently, a gas gain of $\sim1 \times 10^{4}$ and long-term stability 
with a gas gain of $6{\times}10^{3}$ have been realized~\cite{Miuchi1}. 
A position resolution of $120\microm$ was also obtained~\cite{Takeda1}.

Although the electrode structure of the {\upic} reminds that of a Microdot 
chamber (MDOT)~\cite{Biagi1}, its production method is quite different, and 
thus the {\upic} has an advantage of a thick substrate. 
In a MDOT detector, which has a very thin ($\mathrm{< 5 \microm}$) insulator 
layer, the electric field near to the surface of the substrate seems to be 
distorted by the buried readout buses~\cite{Biagi2}. 
On the contrary, this effect is weakened by the thick substrate in the {\upic}.
In this paper, we studied in detail the dependence of the electric field on 
the substrate thickness by a three-dimensional simulation. 
We reveal the operation principle and features of the {\upic}, and discuss 
the optimization of the electrode structure.

\begin{figure}[hp]
\begin{center}
\includegraphics[width=0.7\linewidth]{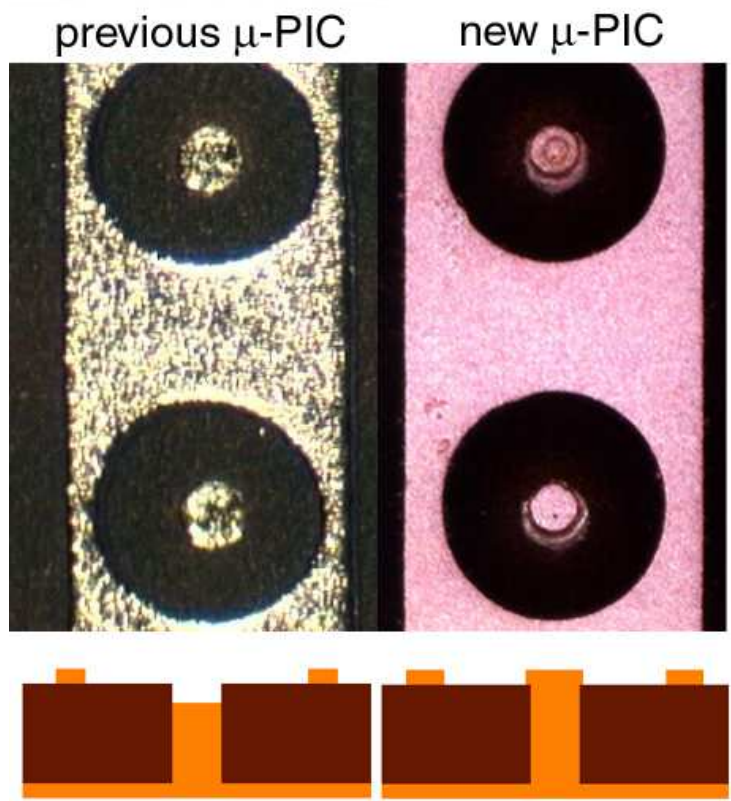}
\caption{Microscopic photograph and schematic cross section of the previous
{\upic} and the new {\upic}. The new {\upic} has higher anode pillars than 
the previous {\upic}.}
\label{fig:1-5c-top.eps}
\end{center}
\end{figure}

\begin{figure}[hp]
 \begin{minipage}{0.5\hsize}
  \begin{center}
  \includegraphics[width=0.95\linewidth]{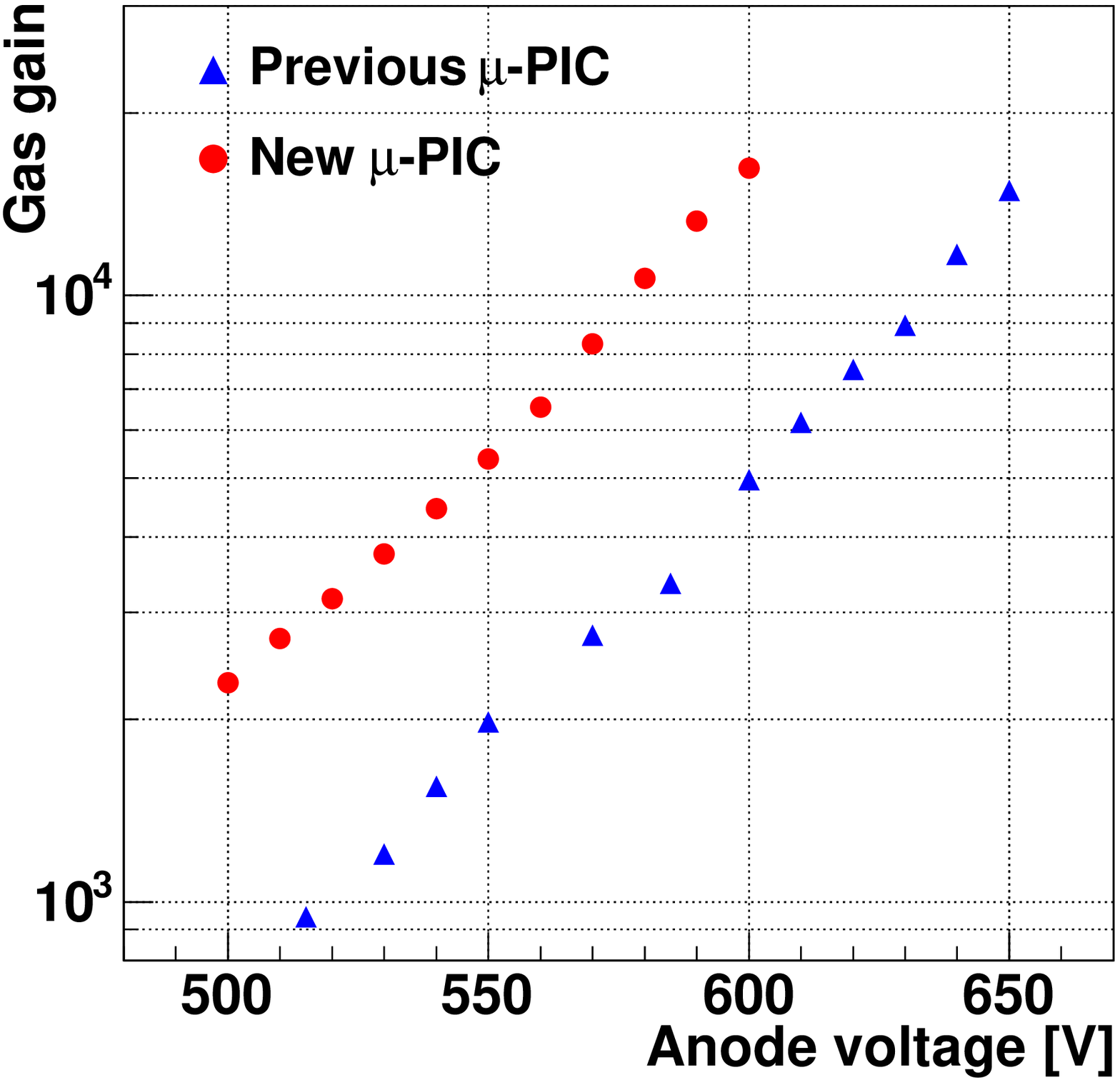}\\
  (a)\\
  \end{center}
 \end{minipage}
 \hfill
 \begin{minipage}{0.5\hsize}
  \begin{center}
  \includegraphics[width=0.95\linewidth]{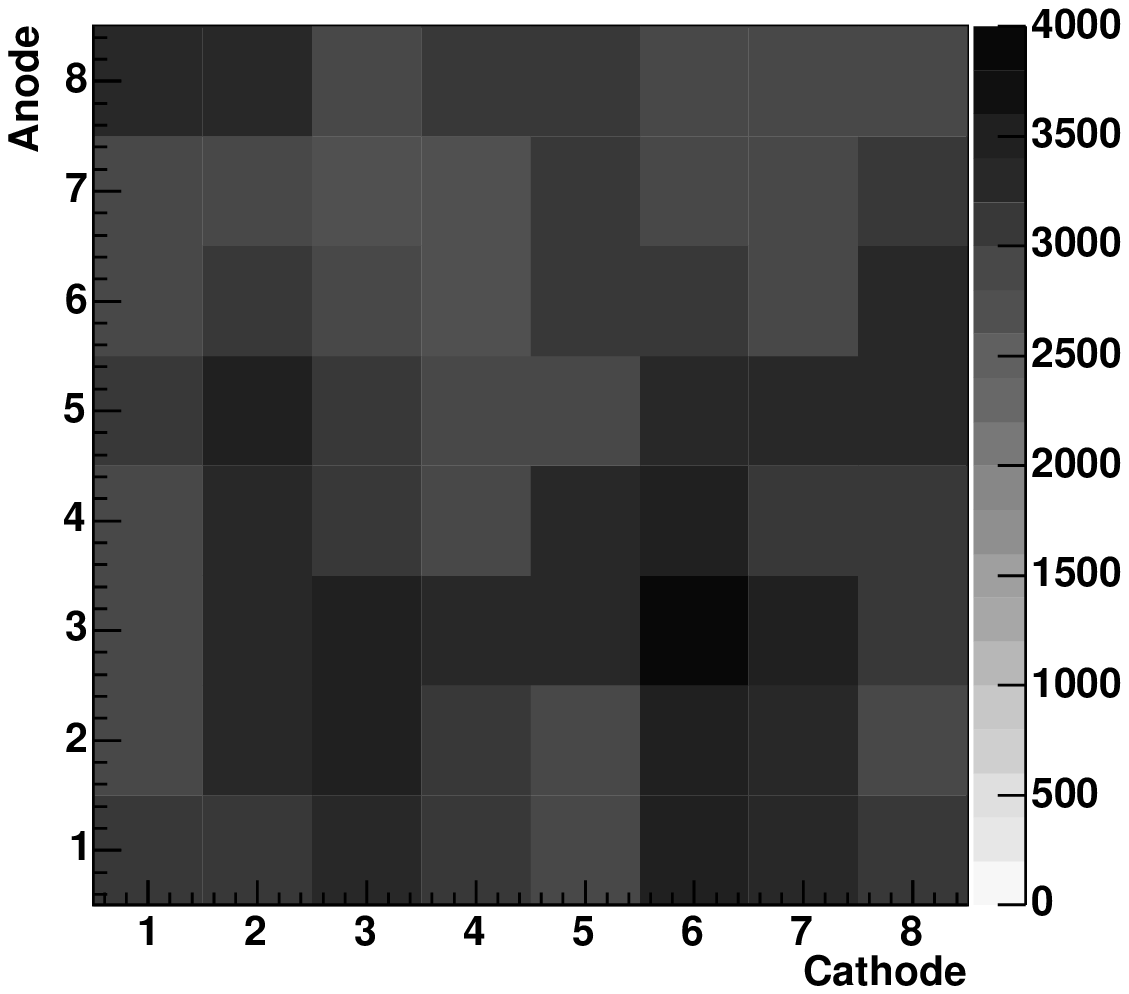}\\
  (b)\\
  \end{center}
 \end{minipage}
 \caption{(a) Gain curve of the new {\upic} (circle) and the previous {\upic}
         (triangle). (b) Map of the measured gas gains. The detection area is
         divided into $8 \times 8$ regions. }
 \label{fig:gain_hikaku2.eps}
\end{figure}

In our previous simulation work~\cite{Bouianov1}, 
we found that the gas gain was degraded because the top of the 
anode pillars are below the substrate; $i.e.$, 
the collection efficiency of primary electrons was about 30{\%}.
With this previous structure, two thirds of the drift electrons ended on 
the substrate, only the rest contributing the observable gas gain.
In addition, these electrons attached on the surface 
possibly caused discharges. 
In order to solve these problems, we have developed a {\upic} with a new 
manufacturing technology~\cite{Nagayoshi2}. 
The anode electrodes of this new type {\upic} are uniformly formed 
to $\mathrm{10\microm}$ above the substrate surface. 
Microscopic photographs and schematic cross sections of the previous and this 
new {\upic}s are shown in Fig.~\ref{fig:1-5c-top.eps}. 
The measured gas gain was 
three times higher than that of the previous {\upic}, as shown in 
Fig.~\ref{fig:gain_hikaku2.eps}(a), which was consistent with the simulation 
results~\cite{Bouianov1}.
Detailed performance studies of this new {\upic} is reported 
elsewhere~\cite{Takeda1}. 
This result shows that the three-dimensional simulation is a reliable method to
study the geometrical structures of the {\upic}.
Therefore, we study the operation principles quantitatively and investigated 
the optimum structure of the {\upic} with a three-dimensional simulation.

\section{Simulation method}
\label{sec:method}
\subsection{Codes}
\label{sec:structure_of_uPIC}
Simulation studies were carried out using Maxwell 3-D field simulator 
(Maxwell)~\cite{Maxwell} and Garfield~\cite{Garfield}. 
The electromagnetic field in a three-dimensional structure was calculated 
using a finite element method (FEM) by Maxwell. 
Maps of the electric field, voltage, and dielectric constant were generated 
by the Maxwell for the unit cell of the {\upic},
which is represented as a mesh consisting of fine tetrahedrons.
A unit cell is 400$\times$400$\microm^2$ in size and consists of a polyimide 
substrate with a relative dielectric constant of 3.5 and a pair of anode 
and cathode electrodes.

These maps were then fed to Garfield, and the dynamics of the drift electrons 
were simulated.
Primary electrons were uniformly generated in the unit area 400$\microm$ 
above the substrate, and the drift paths of electrons were calculated 
by the Monte Carlo method.
The absolute multiplication factor $M_{\mathrm{abs}}$ was calculated 
as an exponentiated integral of the first Townsend coefficient along
the drift path,
\begin{equation}
M_{\mathrm{abs}} = \exp \! \left( \int \! \alpha \, \mathrm{d}z \right),
\label{eq:gas_gain}
\end{equation}
where $\alpha$ is the first Townsend coefficient.
The Townsend coefficient of the gas mixture was calculated using the 
Magboltz~\cite{Magboltz} program.
Usually, some of the electrons do not reach the anodes, 
and hence the observed, or ''effective'' multiplication 
factor $M_{\mathrm{eff}}$ is smaller than $M_{\mathrm{abs}}$.
We define $R_{\mathrm{eff}}$, an electron collection efficiency by 
\begin{equation}
M_{\mathrm{eff}} = R_{\mathrm{eff}}M_{\mathrm{abs}}.
\label{eq:eff_gas_gain}
\end{equation}
In the simulation, we estimated $R_{\mathrm{eff}}$ by 
$R_{\mathrm{eff}} = N_{\mathrm{col}}/N_{\mathrm{tot}}$, where 
$N_{\mathrm{col}}$ is the number of primary electrons collected by the anode, 
and $N_{\mathrm{tot}}$ is the total number of primary electrons 
that reached the detector surface. 
Because the observed gas gain, or the detector response, 
is determined by $M_{\mathrm{eff}}$, 
we focus on $M_{\mathrm{eff}}$ in the following discussion.

\subsection{Parameters}
\label{sec:basic_parameters}
The geometrical parameters of the {\upic} used for the simulation 
are listed below, where $h$ is the height from the surface of the substrate.
\begin{enumerate}
\item Height of the anode electrode (fixed; at $h=10\microm$),
\item Width of the rear side anode strip (fixed; at $200\microm$),
\item Width of the cathode strip (fixed; $314\microm$),
\item Diameter of the cathode openings (fixed; $260\microm$),
\item Substrate thickness~$t$ 
      (variable; $\mathrm{default = 100\microm}$),
\item Diameter of the anode electrode~$d$ 
      (variable; $\mathrm{default = 50\microm}$).
\end{enumerate}

\begin{figure}[hp]
\begin{center}
\includegraphics[width=0.7\linewidth]{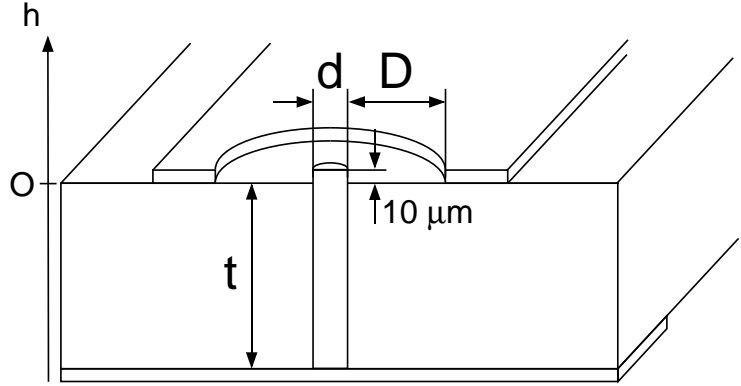}
\caption{Description of the geometrical parameters of the {\upic}.}
\label{fig:danmen_5.eps}
\end{center}
\end{figure}

We examined the $M_{\mathrm{eff}}$ dependences on $t$ and $d$, 
with all other parameters fixed. 
These geometrical parameters of the {\upic} are described in 
Fig.~\ref{fig:danmen_5.eps}. 
The anode-cathode voltage was 600V and the drift electric field 
was $\mathrm{1kV/cm}$. 
The gas mixture was Ar/$\mathrm{C_{2}H_{6}}$ (80/20).

\section{Simulation results}
\label{sec:result}
\subsection{$M_{\mathrm{eff}}$ dependence on the substrate thickness}
\label{sec:thickness}
Some pixel type detectors were made by the IC technology
and hence had a very thin insulator between metal layers (several $\microm$). 
Therefore, the electric field in the drift region near the surface of 
the substrate is affected by the readout strips buried in the substrate. 
Some of the drift electrons would be directed toward the readout buses, 
rather than the anode electrodes. 
This effect would be observed as a degradation of the gas gain, or small 
$R_{\mathrm{eff}}$ values. 
Furthermore, the accumulated charge on the substrate might cause discharges. 
There are two approaches to avoid this problem. 
The first is to make a narrow-pitch detectors~\cite{Han1}. 
Actually, a narrow pitch ($< 100\microm$) MDOT detector was reported to have 
overcome this problem. 
Because discharges could be problematic with the small anode-cathode distance
for a large area detector having huge number of pixels,
we adopted the other approach; a thick substrate.
This approach was only realized with the PCB technology.

\begin{figure}[hp]
 \begin{minipage}{0.5\hsize}
  \begin{center}
  \includegraphics[width=0.95\linewidth]{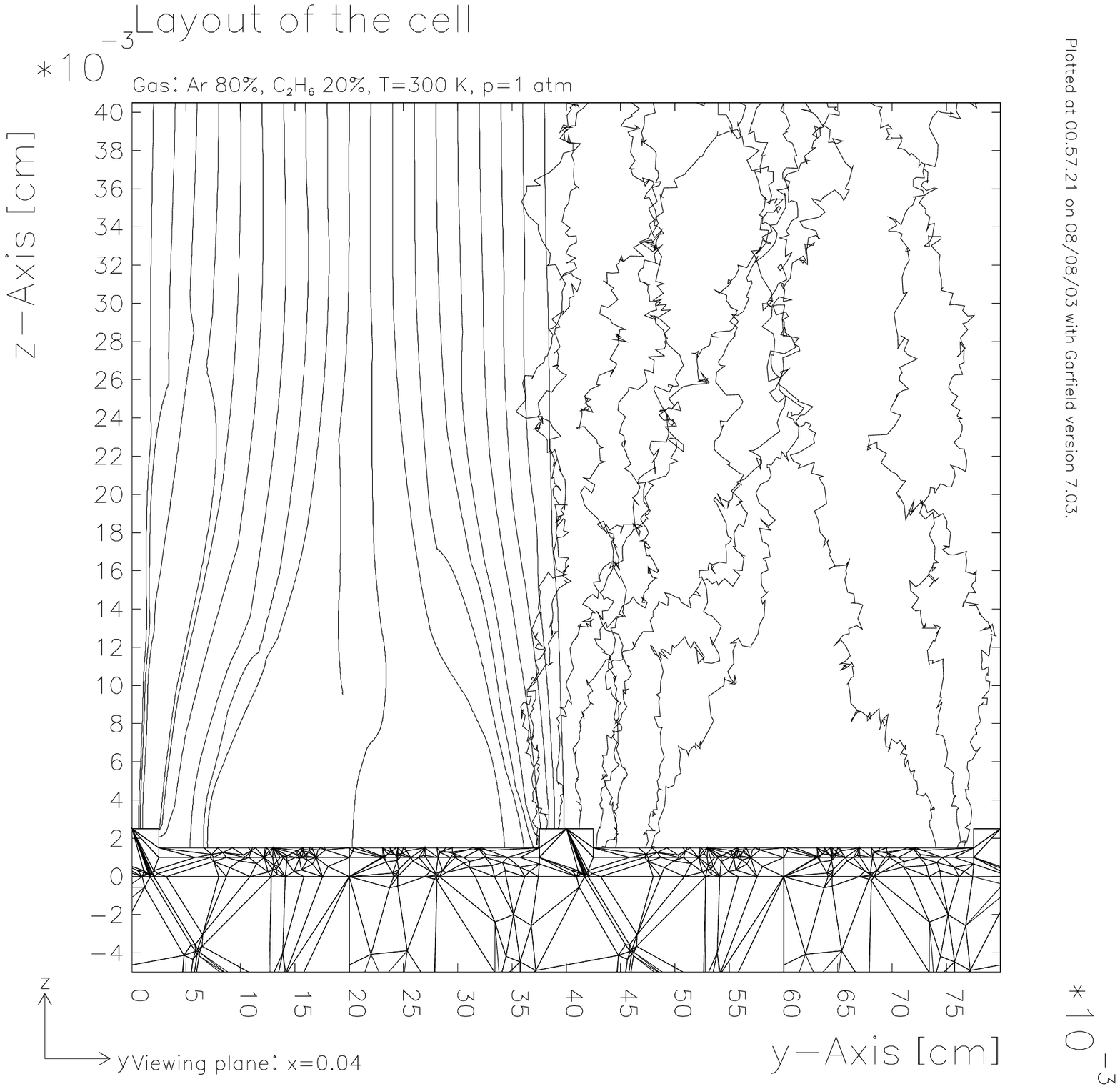}\\
  (a)\\
  \end{center}
 \end{minipage}
 \hfill
 \begin{minipage}{0.5\hsize}
  \begin{center}
  \includegraphics[width=0.95\linewidth]{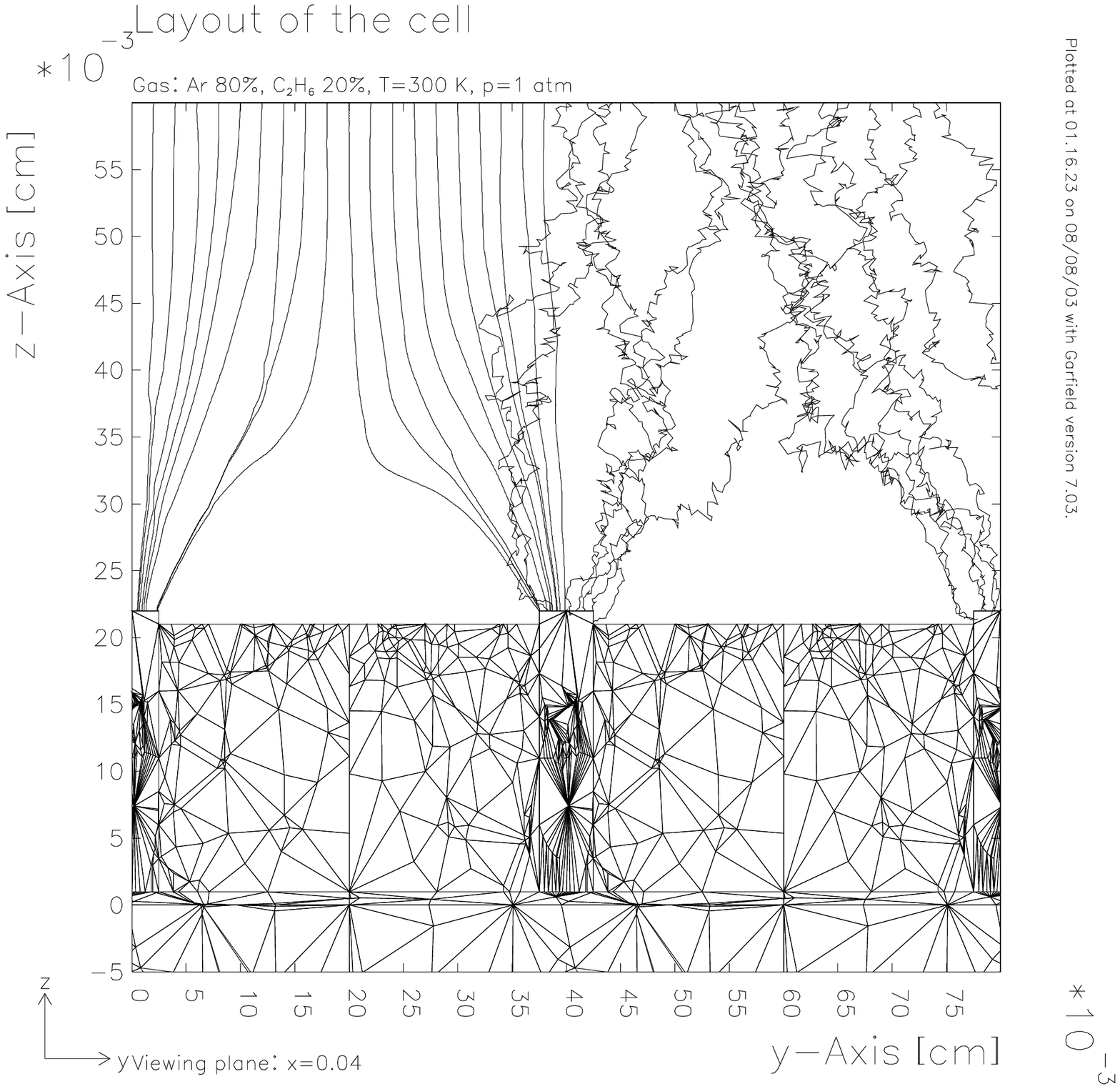}\\
  (b)\\
  \end{center}
 \end{minipage}
 \caption{Simulated electron drift lines of (a)$t=5\microm$,
          and (b) $t=200\microm$. 
          In both figures, the drift lines drawn in the left half were 
          calculated by the Runge Kutta Fehlberg method, and those in the right
          half were simulated by the Monte Carlo method.}
 \label{fig:drift_line}
\end{figure}

\begin{figure}[hp]
 \begin{minipage}{0.5\hsize}
  \begin{center}
  \includegraphics[width=0.95\linewidth]{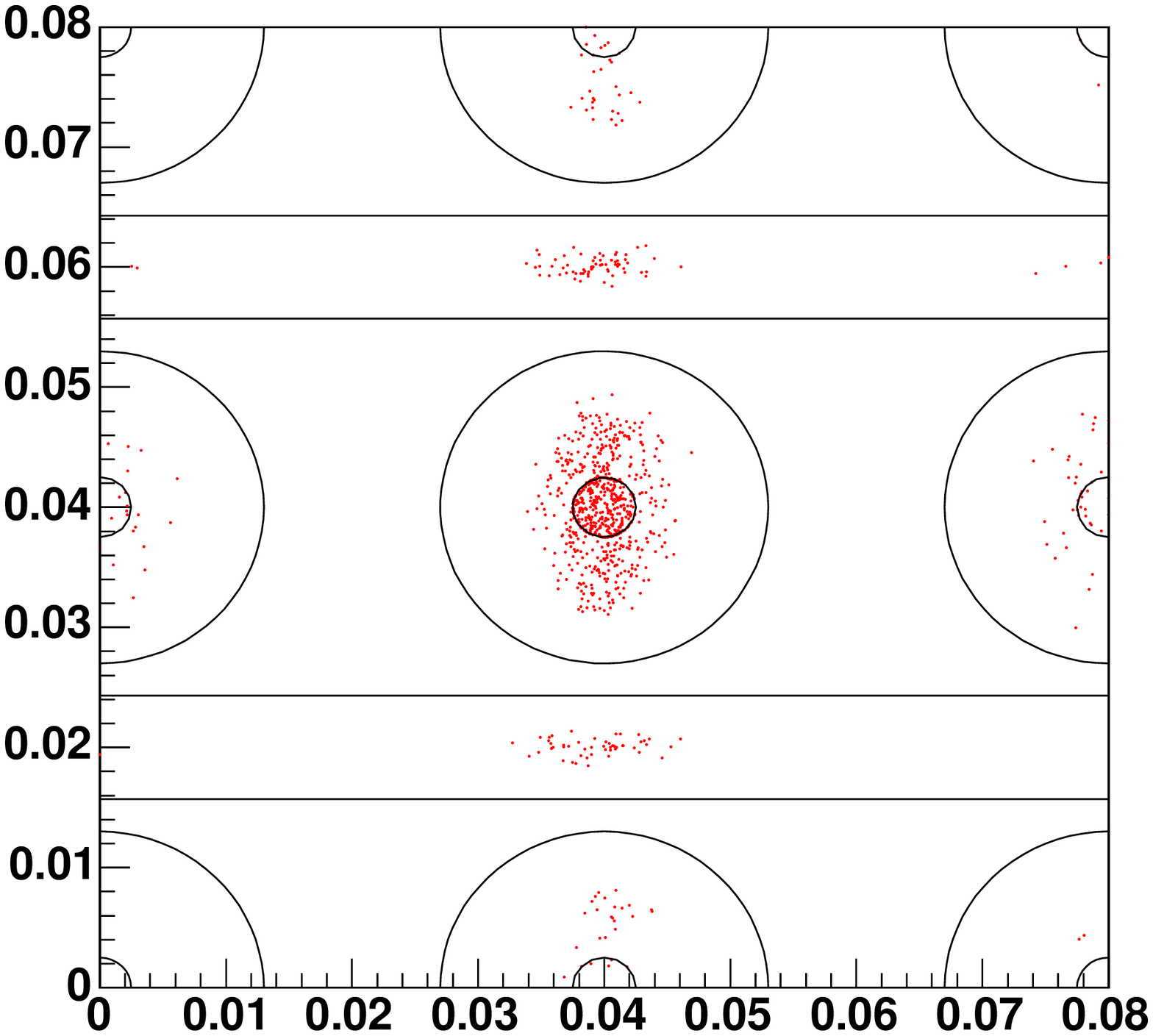}\\
  (a)\\
  \end{center}
 \end{minipage}
 \hfill
 \begin{minipage}{0.5\hsize}
  \begin{center}
  \includegraphics[width=0.95\linewidth]{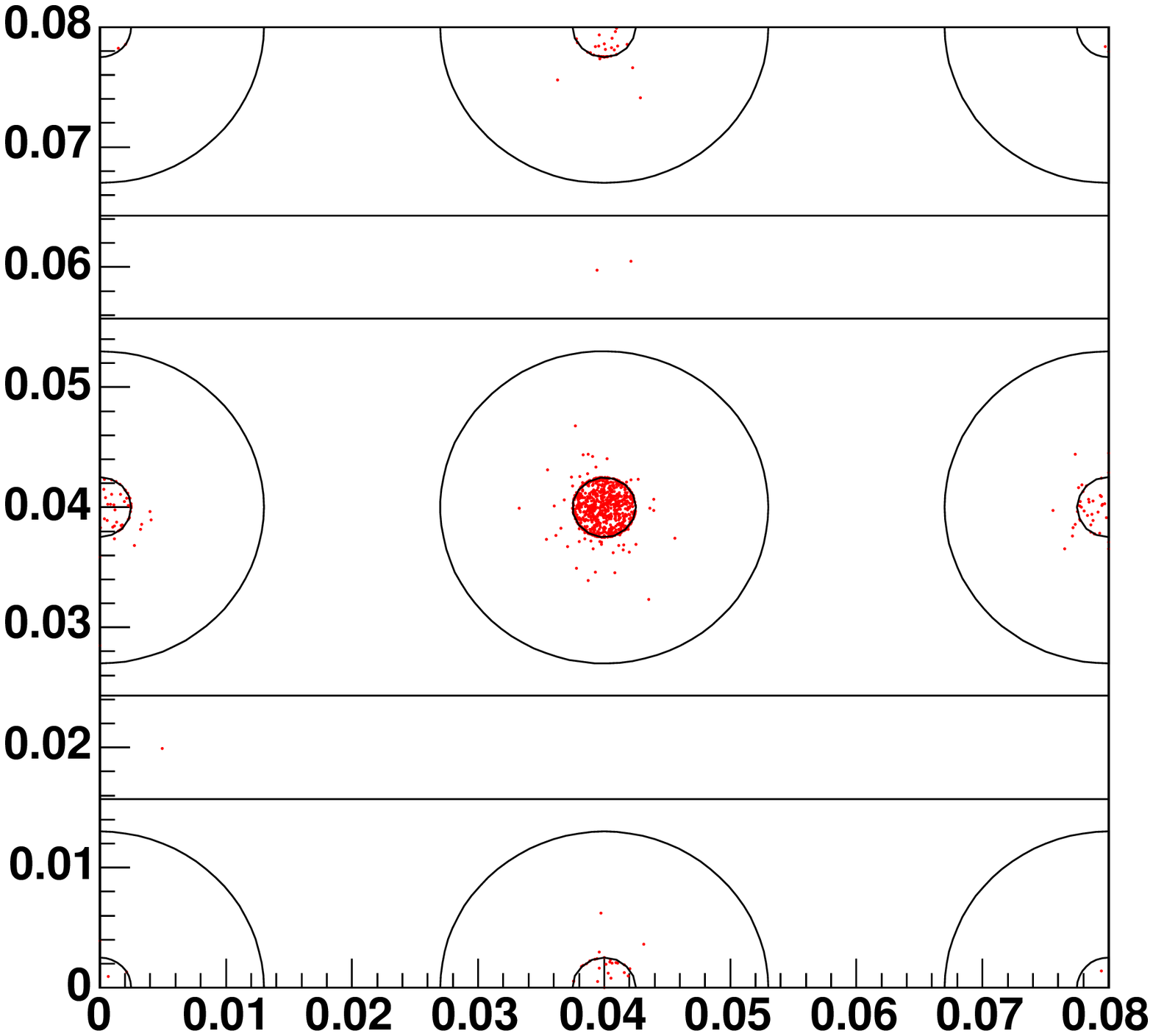}\\
  (b)\\
  \end{center}
 \end{minipage}
 \caption{Terminal points of electrons on the electrode plane of the {\upic} 
of (a) $t=5\microm$, and (b) $t=200\microm$. 
1000 events are shown in both cases.}
 \label{fig:endpoint_kiban}
\end{figure}

\begin{figure}[hp]
\begin{center}
  \includegraphics[width=0.7\linewidth]{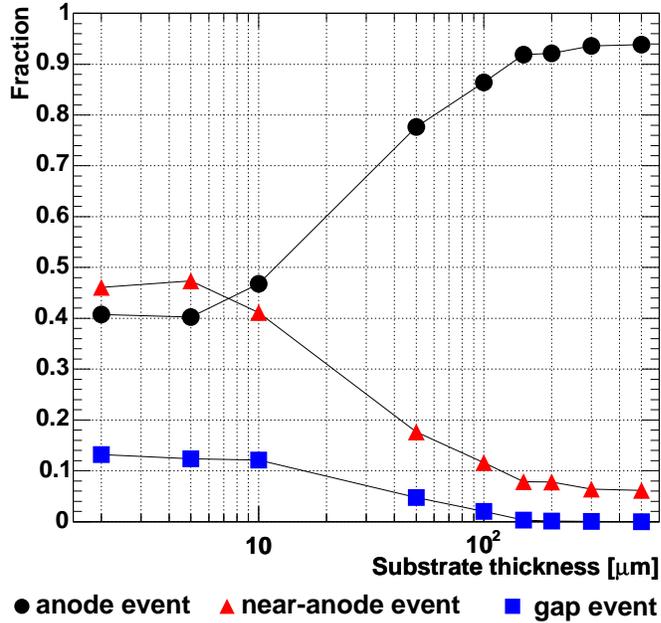}
   \caption{Electron collection efficiency dependence on $t$ (circles). 
	For reference, fractions of electrons that are 
            terminated on the substrate in the cathode openings and the gaps 
            between the cathodes are shown by the 
	triangles and squares, respectively.}
  \label{fig:elec_effic_kiban.eps}
\end{center}
\end{figure}

Figs.~\ref{fig:drift_line} and~\ref{fig:endpoint_kiban} show the electron drift
lines and the distribution of terminal points of the drift electrons, where 
the results of the thin ($t=5\microm$) and thick 
($t=200\microm$) substrates are shown in (a)s and (b)s, respectively. 
For $t=5\microm$, some of the drift lines end at the substrate 
(Fig. ~\ref{fig:drift_line} (a)) and the resultant electrons are terminated 
on the substrate (Fig.~\ref{fig:endpoint_kiban} (a)).
It was obvious that the buried anode readout buses distorted 
the drift electric field for $t=5\microm$, 
because the electron end points are distributed above the anode strips 
which are placed orthogonally to the cathode strips. 
On the other hand, most electrons are collected to the anode electrodes for
$t=200\microm$. 
Fig.~\ref{fig:elec_effic_kiban.eps} shows $R_{\rm eff}$ dependence on $t$.
In the case of a thin ($t< 10\microm$) substrate, 
$R_{\rm eff}$ was found to be less than 0.5,
while that with a thick ($t> 100\microm$) substrate exceeds 0.8.
These results indicate that the {\upic} with $100\microm$ thick substrate 
already has a structure close to the optimal one in the view point 
of $R_{\rm eff}$ dependence on the substrate thickness.

\begin{figure}[hp]
 \begin{minipage}{0.5\hsize}
  \begin{center}
  \includegraphics[width=0.95\linewidth]{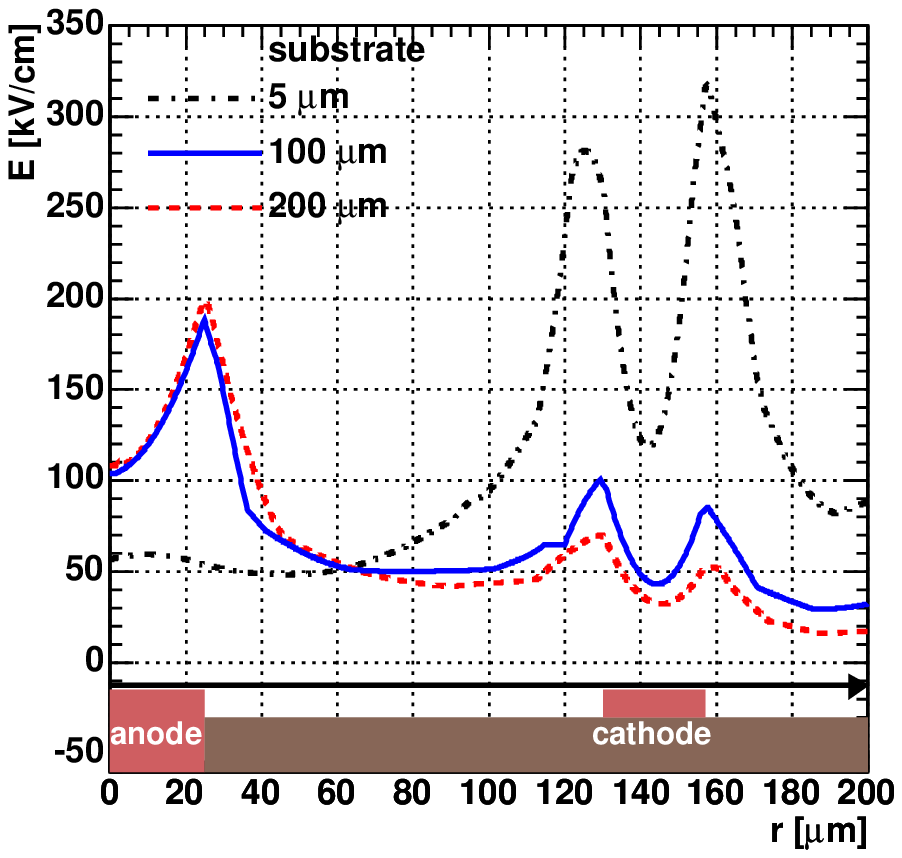}\\
  (a)\\
  \end{center}
 \end{minipage}
 \hfill
 \begin{minipage}{0.5\hsize}
  \begin{center}
  \includegraphics[width=0.95\linewidth]{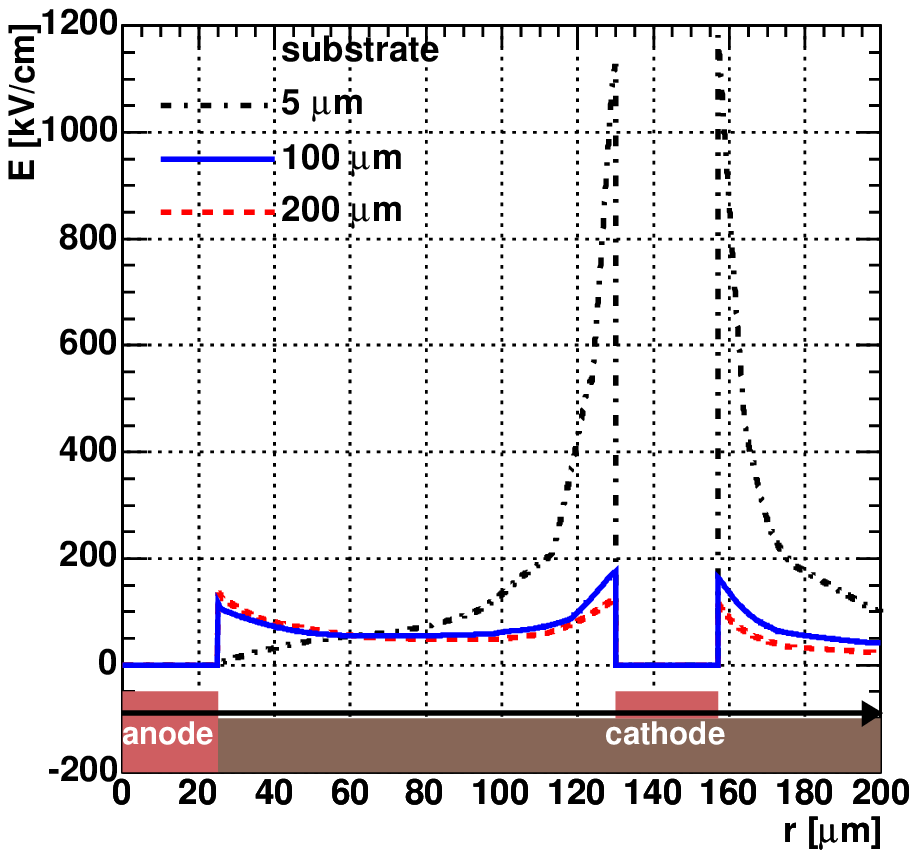}\\
  (b)
  \end{center}
 \end{minipage}
 \caption{Strength of the electric field as a function of the distance from the
          anode center: 
          (a) the $r$ axis is located $1\microm$ above the top of the
          anode electrodes.
          (b) the $r$ axis is located $1\microm$ above the substrate.}
 \label{fig:field_r_substrate}
\end{figure}

\begin{figure}[hp]
\begin{center}
\includegraphics[width=0.7\linewidth]{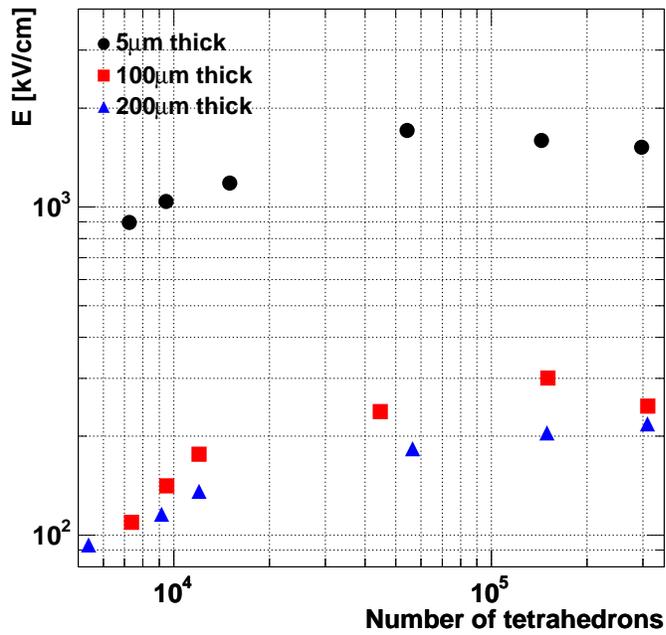}
\caption{The mesh dependence of the electric field at the cathode edge for
the {\upic}s with several substrate thickness.}
\label{fig:e_field_kiban_mesh.eps}
\end{center}
\end{figure}

Another parameter which determines $M_{\mathrm{eff}}$ is $M_{\mathrm{abs}}$. 
Because $M_{\mathrm{abs}}$ strongly depends on the electric field around 
the anode electrodes, $M_{\mathrm{abs}}$ also depends on $t$. 
Fig.~\ref{fig:field_r_substrate} (a) shows the electric fields as a function of
the distance $r$ from the anode center at $h=11\microm$ 
($1\microm$ above the top of an anode).
With a thin ($t{\Ltsim}5\microm$) substrate,  
the electric field averaged over the top area of the anode electrodes 
(hereafter $E_{\mathrm a}$) reaches only 60kV/cm, while that with a thick 
($t>100\microm$) substrate exceeds 150kV/cm.
For a thin substrate, the electron drift is affected 
by the rear side anode strip. 
Then large fraction of the drift electrons are not focused on the anode pixels.
On the other hand, the electrode structure is similar to a wire chamber 
for a thick substrate. 
Most electrons are directed toward the anode pixels and collected on the 
anode.
Thus a large $M_{\mathrm{abs}}$ is attainable for a thicker substate.
The electric fields near the surface of the substrate 
($h=1\microm$) are also shown in Fig.~\ref{fig:field_r_substrate} (b) 
as a function of $r$, where the $r$ axis is located $1\microm$ above 
the substrate.
The electric field averaged along the circumference of the cathode ring
(hereafter $E_{\mathrm c}$) is found to be 10 times stronger for a thin 
($t=5\microm$) substrate than that for a thick ($t>100\microm$) substrate. 
When we discuss the electric field around sharp points such as the 
cathode edge, the accuracy of the simulation has to be carefully considered.
The detector structure and drift space consist of fine tetrahedrons in the FEM
calculation. 
Then the number of tetrahedrons $N_{\mathrm{tetra}}$ may represents the 
accuracy of the calculation.
In this simulation, a unit cell includes 12000-16000 tetrahedrons.
The mesh size is several $\microm$ which is smaller enough than the anode
diameter.
Since this $N_{\mathrm{tetra}}$ does not seem to be very large for precise 
calculation, the validity of this accuracy was checked as follows.
The electric fields at the cathode edge for several substrate thickness were 
calculated as functions of the number of tetrahedrons in the unit 
cell (Fig.~\ref{fig:e_field_kiban_mesh.eps}).
While the mesh size is relatively large ($N_{\mathrm{tetra}} < 10000$), 
the electric field at the cathode edge gradually increases as the number of
tetrahedrons.
For the larger number of tetrahedrons ($N_{\mathrm{tetra}} > 10000$), 
the resultant electric fields reach their plateau and can be regarded as 
constant within the error of 15{\%} in RMS.
Thus we used the $N_{\mathrm{tetra}}$ = 12000-16000, which are reasonable 
numbers to save the CPU time.

\begin{figure}[hp]
\begin{center}
\includegraphics[width=0.7\linewidth]{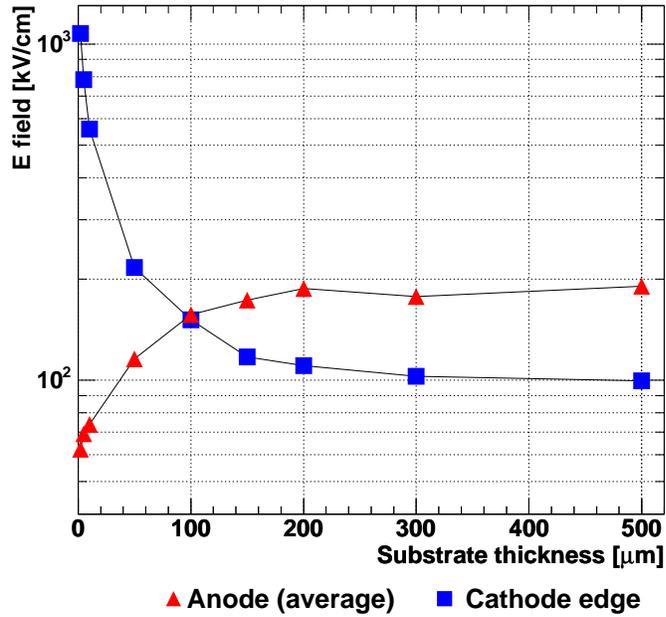}
\caption{Strength of the electric field at an anode pixel 
         (triangle), and near to cathode edge (square) as functions of 
         the substrate thickness. The anode voltage was 600V for all 
         calculations.}
\label{fig:field_kiban3.eps}
\end{center}
\end{figure}

The Townsend coefficient for a thin substrate becomes larger near 
the cathode edge, which is likely to cause discharges due to the emission of 
electrons from the cathode electrode. 
This is because the distance between the cathode edge and
the anode readout bus ($t$) is much shorter than the distance between the
anode pillar and the cathode edge ($D$).
On the other hand, if $t$ is larger than $D$, 
the effect of the readout bus is efficiently shielded by the substrate, 
and the $E_{\mathrm c}$ was weaker than the $E_{\mathrm a}$. 
$E_{\mathrm a}$ and $E_{\mathrm c}$ for various $t$ values are 
shown in Fig.~\ref{fig:field_kiban3.eps}.
It is intuitively expected that $E_{\mathrm c}$ should be at least weaker 
than $E_{\mathrm a}$, and the mechanism of discharges will be discussed 
in later section. 
Because $E_{\mathrm a}$ is comparable to $E_{\mathrm c}$ with 
$t\sim 100\microm$, 
thickness of the substrate of the present {\upic} is thought to satisfy 
the minimum point, though it is not still sufficient.
Thicker substrate ($t \Gtsim 150\microm$) would help to realize 
larger $M_{\mathrm{abs}}$ with the {\upic}.

\begin{figure}[hp]
\begin{center}
\includegraphics[width=0.7\linewidth]{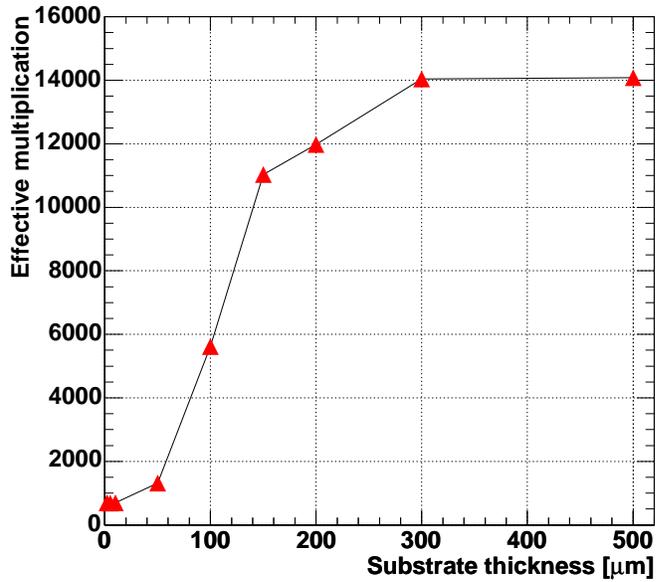}
\caption{Calculated effective multiplication $M_{\mathrm{eff}}$ as 
         a function of the substrate thickness.} 
\label{fig:gasgain_kiban.eps}
\end{center}
\end{figure}

As the consequence of these two studies, thicker substrate is thought to bring 
larger $M_{\mathrm{eff}}$. 
Fig.~\ref{fig:gasgain_kiban.eps} shows $M_{\mathrm{eff}}$ as a function of $t$.
With a substrate thicker than $150\microm$, 
$M_{\mathrm{eff}}$ is expected to be twice larger than the current {\upic}.

\subsection{$M_{\mathrm{eff}}$ dependence on the anode diameter}
\label{sec:diameter}
\begin{figure}[hp]
\begin{minipage}{0.5\hsize}
\begin{center}
\includegraphics[width=0.95\linewidth]{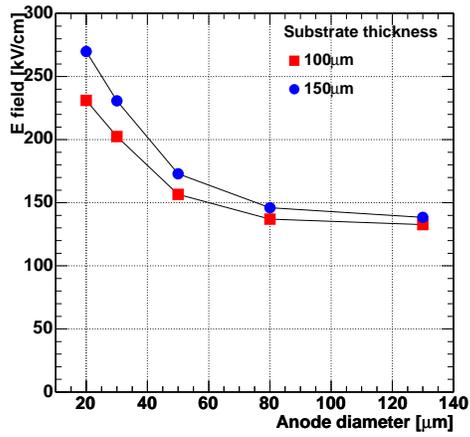}\\
(a)\\
\end{center}
\end{minipage}
\hfill
\begin{minipage}{0.5\hsize}
\begin{center}
\includegraphics[width=0.95\linewidth]{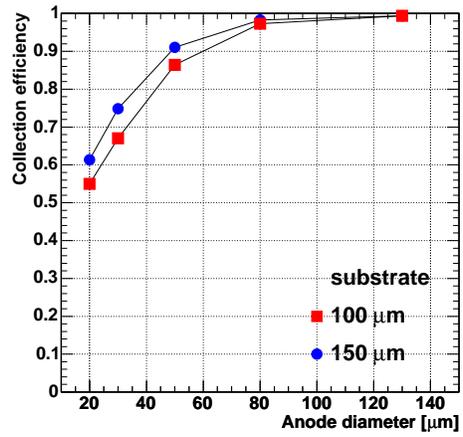}\\
(b)\\
\end{center}
\end{minipage}
\begin{center}
\caption{Electric field near the anode top (a) and electron collection
         efficiency (b) as functions of the anode diameter.}
\label{fig:field_effic}
\end{center}
\end{figure}

\begin{figure}[hp]
\begin{center}
\includegraphics[width=0.7\linewidth]{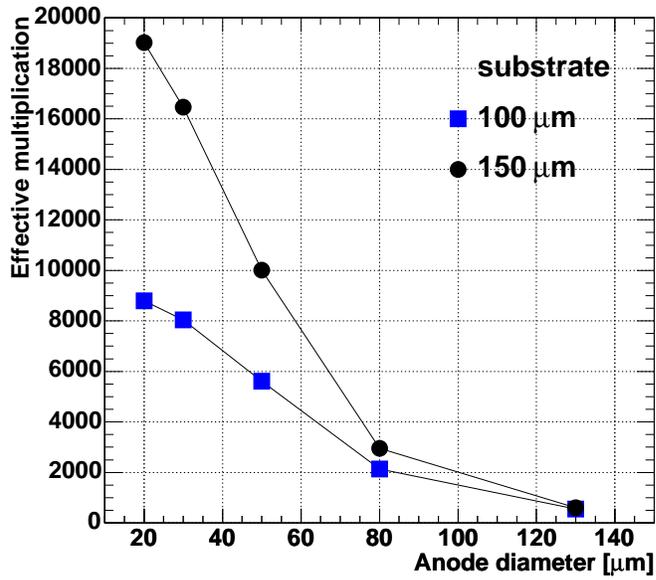}
\caption{Dependence of the effective multiplication factor as a function of the
anode diameter.}
\label{fig:gas_gain_diam3.eps}
\end{center}
\end{figure}

The dependences of the $R_{\mathrm{eff}}$ and $M_{\mathrm{abs}}$ 
on the anode diameter ($d$) were also studied.
Although the current PCB technology can not manufacture an anode electrode
with a diameter of smaller than $50\microm$ for a thick substrate, 
it is worth investigating the optimum diameter.
In this simulation, the anode diameter was varied from $20\microm$ to 
$130\microm$ and the electron drift and multiplication were calculated.
Fig.~\ref{fig:field_effic} (a) shows $E_{\mathrm a}$ as a function of $d$. 
Obviously, $E_{\mathrm{a}}$ decreases with increasing $d$.
On the other hand, large $R_{\mathrm{eff}}$ is expected for the 
large $d$, as shown in Fig.~\ref{fig:field_effic} (b). 
Fig.~\ref{fig:gas_gain_diam3.eps} shows the effective multiplication factor
$M_{\mathrm{eff}}$ as a function of $d$. 
As is well known, this result shows that a smaller $d$
20-30$\microm$ has a larger gain; however, it also indicates that a sufficient 
multiplication is expected for an $ d\sim 50\microm$.

\section{Detector optimization}
\label{sec:discussion}
In the previous section, we showed quantitatively that a thick substrate and a
small diameter the of anode electrode 
would help to improve the performance of the {\upic}. 
Because a smaller anode electrode is technically unreachable in the current
{\upic} fabrication method,
we discuss the maximum $M_{\mathrm{eff}}$ of the {\upic} with a 
thick substrate considering the maximum anode voltage for stable operation
for a $50\microm \phi$ anode.

As indicated in earlier works about MSGCs~\cite{Nagae1,Pestov1,Beckers1,Oed1},
one of the dominant mechanism of discharges in micropattern gas detectors is a 
consequence of the ejection of electrons from the cathode edge by spontaneous 
field emission.
At a normal operation condition of MSGCs, 
the electric field near the cathode edge exceeds 100kV/cm, 
which is high enough for gas avalanche~\cite{Beckers1}.
Therefore, electrons released from the cathode electrodes undergo the gas 
multiplication near both the cathode and anode electrodes, 
while the primary electrons have gas multiplication only near the 
anode electrode.
We can assume that the electric field near the cathode governs the discharge 
process and ``stable operation'' can be judged by the cathode field.
It is natural to think this discharge mechanism can be applied to {\upic} 
and we can determine the maximum anode voltage by $E_{\mathrm{c}}$.
As a result, the maximum $M_{\mathrm{eff}}$ is determined by $E_{\mathrm{c}}$.

\begin{figure}[hp]
\begin{center}
\includegraphics[width=0.7\linewidth]{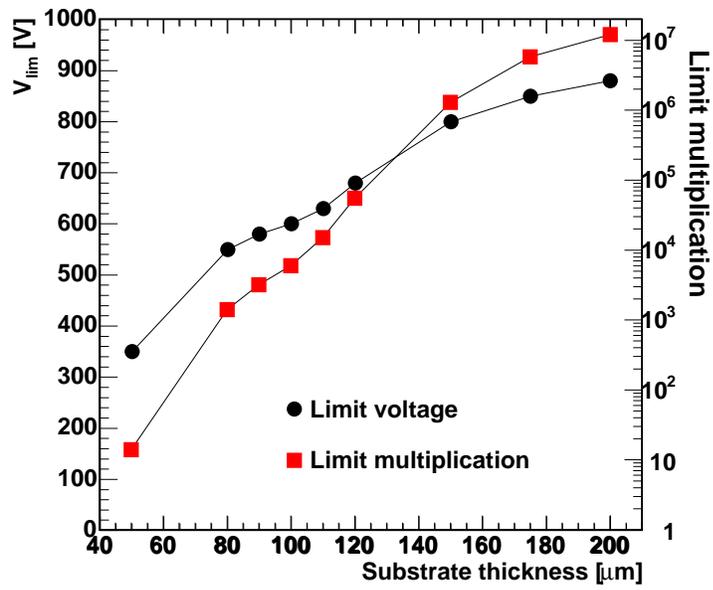}
\caption{Dependence of the limit voltage (filled circle) and the corresponding 
         achievable gas gain (filled square) as a function of the 
         \textit{Thickness-Distance Ratio}.}
\label{fig:vmax_kiban2.eps}
\end{center}
\end{figure}

From the experimental results~\cite{Miuchi1}, 
we knew that the maximum allowable anode voltage was about 630V for 
$\mathrm{Ar/C_{2}H_{6}}$ (80/20) gas mixture, 
which corresponds to the maximum $E_{\mathrm c}$ of 200kV/cm.
We use 200kV/cm as the limit of $E_{\mathrm c}$.
As mentioned in the previous sections, $E_{\mathrm c}$ 
becomes weaker for a thicker substrate with the same anode voltage. 
In other words, the corresponding allowable voltage $V_{\mathrm{lim}}$ becomes 
higher for a larger $t$.
In Fig.~\ref{fig:vmax_kiban2.eps}, $V_{\mathrm{lim}}$ and corresponding 
$M_{\mathrm{eff}}$ are shown as functions of the substrate thickness.
For a thicker substrate of $t > 150\microm$, 
maximum anode voltage exceeds 800V, which the maximum gas multiplication 
for that voltage is higher than $10^{6}$, as shown 
in Fig.~\ref{fig:vmax_kiban2.eps}. 
Taking into account the other factors, 
such as the non-uniformity of the shape of electrode,
the gas gain might achieve $10^{5}$ for a $150\microm$ thick substrate.

\section{Summary}
\label{sec:summary}
A three-dimensional simulation of the electron drift and gas multiplication in 
a {\upic} was performed for optimizing the electrode structure using Maxwell 
3-D field simulator and Garfield software. 
We found that the substrate thickness is a key factor for achieving both 
a high gas gain and stable operation. 
The electron collection efficiency $R_{\mathrm{eff}}$ and the gas 
multiplication factor $M_{\mathrm{eff}}$ became higher for a thicker substrate.
In order to obtain sufficient gas gain for detecting minimum ionization 
particles, the substrate thickness of more than $150\microm$ is required. 
Furthermore, a thick substrate also has an advantage for discharge suppression.
The electric field near the cathode edge is weaker for a thicker substrate.
The dependence of gas multiplication on the anode diameter was also 
simulated. 

We estimated a limit voltage $V_{\mathrm{lim}}$ as a function of the substrate
thickness based on experimental data, and found that $V_{\mathrm{lim}}$ would 
reach 800V when thickness is $\sim 150\microm$. 
In addition, a quite high gas gain of $\sim 10^{5}$ would be possible. 
We aim to achieve a thickness of $\Gtsim 150\microm$ in the next improvement of
the {\upic}.

\section*{Acknowledgments}
This work is supported by a Grant-in-Aid for the 21st Century COE ``Center for 
Diversity and Universality in Physics'', and a Grant-in-Aid in Scientific 
Research of the Japan Ministry of Education, Culture, Science, Sports and 
Technology, and ``Ground Research Announcement for Space Utilization'' promoted
by Japan Space Forum.
Also this work is partially supported by JAERI and KEK.
T. N., R. O., and M. U. acknowledge the receipt of JSPS Research Fellowship.

\clearpage


\clearpage

\clearpage

\end{document}